\documentclass{article}%
\usepackage{amsmath}
\usepackage{amssymb}
\usepackage{amsfonts}
\usepackage{graphicx}%
\providecommand{\U}[1]{\protect\rule{.1in}{.1in}}

\begin{document}

\title{Derivation of the Boltzmann equation with no "molecular chaos"-type approximation}
\author{Victor F. Los, V.G. Baryakhtar Institute of Magnetism of
\and Nat. Acad. of Sci. of Ukraine, 36b Vernadsky Blvd., 03142, Kyiv, Ukraine \ \ }

\begin{abstract}
The paper resolves the problem of the derivation of a completely closed
evolution equation for $s$-particle distribution function $F_{s}(t)$ ($s\leq
N$) from the Liouville equation for $N\gg1$-particle distribution function
$F_{N}(t)$\ with arbitrary initial condition $F_{N}(0)$ and without any use of
the "molecular chaos" type approximation. The initial correlations are
accounted for in this equation in the kernel governing the evolution of
$F_{s}(t)$ via the special projection operator which exactly transforms the
inhomogeneous Nakajima-Zwanzig Generalized Master Equation (GME) with an
irrelevant initial condition term into the homogenous one. This equation is
further simplified by presenting its kernel in the linear in the particles'
density $n$ approximation. In this approximation the equations for
one-particle $F_{1}(t)$ and two-particle $F_{2}(t)$ distribution functions are
derived. It is shown that the terms describing the influence of initial
correlations in the equation for $F_{1}(t)$ disappear at the large timescale
$t\thicksim t_{rel}\gg t_{cor}$ ($t_{cor}$ is a short correlation time as
compared to a relaxation time $t_{rel}$ of $F_{1}(t)$) resulting in the linear
Boltzmann equation. This equation can be presented as the nonlinear Boltzmann
equation in the time interval $t_{cor}\ll t\ll t_{rel}$. At $t_{rel}%
\rightarrow\infty$ (mean free path $l\rightarrow\infty$) the Boltzmann
equation holds for all finite times $t\gg t_{cor}$.

\end{abstract}
\maketitle

\section{Introduction}

One of the basic tasks of the nonequilibrium statistical mechanics remains
deriving the appropriate evolution equations for the measurable values
(statistical expectations) characterizing a nonequilibrium state of a
many-particle system. The principal question is how to rigorously derive such
irreversible (kinetic or transport) equations from the underlying microscopic
reversible classical or quantum dynamic equations. The kinetic equations are
expected to be completely closed (homogeneous) time-local equations. Several
approaches are usually used to address this problem, commonly starting with
the linear Liouville equation for a distribution function of an $N$-particle
($N\gg1$) system under consideration. However, this equation is not tractable
from a practical point of view. For practical purposes it is sufficient to
derive the time evolution of the probability distributions (marginals) for a
much smaller group of $s\ll N$ particles (e.g., the Boltzmann and Landau
equations describe the evolution of a one-particle distribution function).

One approach to achieving this goal leads to the chain of $N$ coupled
first-order differential equations for $s$-particle distribution functions
$F_{s}(t)$ ($1\leq s\leq N$) known as the BBGKY hierarchy (from the names of
Bogoliubov, Born, Green, Kirkwood and Yvon) \cite{Bogoliubov (1946)}. However,
this method faces a difficulty of the approximate decoupling of the BBGKY
chain and obtaining a closed evolution equation for an $s$-particle marginal.
Boltzmann was the first who decoupled a two-particle distribution function
introducing the "molecular chaos" approximation as%

\begin{equation}
F_{2}(\mathbf{r}_{1},\mathbf{v}_{1};\mathbf{r}_{2},\mathbf{v}_{2}%
;t)=F_{1}(\mathbf{r}_{1},\mathbf{v}_{1};t)F_{1}(\mathbf{r}_{2},\mathbf{v}%
_{2};t), \label{1}%
\end{equation}
where $F_{2}(\mathbf{r}_{1},\mathbf{v}_{1};\mathbf{r}_{2},\mathbf{v}_{2};t)$
is the two-particle distribution function ($\mathbf{r}_{i}$,$\mathbf{v}_{i}$
are the coordinate and velocity of the $i$-th particle, correspondingly) and
$F_{1}(\mathbf{r}_{i},\mathbf{v}_{i};t)$ is the one-particle distribution
function, and thereby obtained his famous nonlinear closed equation for
$F_{1}(\mathbf{v}_{1};t)$. Equation (\ref{1}) means that the two given
particles are uncorrelated at any time moment $t$. But if it is even true at
any initial moment of time $t_{0}$, it cannot be so after particles collision,
and, therefore, Eq. (\ref{1}) assumes the "propagation of chaos". The latter
is supposed to be true in the so-called "Boltzmann-Grad limit"
\cite{Grad1,Grad2}. Note, that nonlinearity of the Boltzmann equation for a
one-particle marginal, obtained from the linear Liouville equation, is a
consequence of the "molecular chaos" approximation (\ref{1}) at any time
moment $t$. Among the rigorous results, Lanford's derivation of the Boltzmann
equation (however only on a small timescale) \cite{Lanford (1975)} seems to be
the most relevant result in the mathematical foundation of the kinetic theory.
Unfortunately, the small time (as compared to the relaxation time) restriction
is serious and insufficient for studying the evolution of the system under
consideration, e.g., toward the equilibrium or stationary state.

The more sophisticated approach to decoupling the BBGKY hierarchy was
developed by Bogoliubov \cite{Bogoliubov (1946)}. He suggested the principle
of weakening of initial correlations, which implies that at a sufficiently
large time $t-t_{0}\gg t_{cor}$ ($t_{cor}$ is the correlation time due to
interparticle interaction), all initial correlations (existing at the initial
instant $t_{0}$) are damped and the time-dependence of multiparticle
distribution functions is consequently completely determined by the
time-dependence of a one-particle distribution function (Bogoliubov's ansatz).

In the projection operators approach, one obtains the so-called linear
Generalized Master Equations (GMEs) (see, e.g., \cite{Breuer}) for a relevant
part of an $N$-particle distribution function, which contain the undesired
(irrelevant) inhomogeneous term (a source) containing all $N$-particle initial
correlations. The same equation can be also obtained from the BBGKY chain
(see, e.g., \cite{Balescu1}). In order to obtain the completely closed
(homogeneous) linear equation for the reduced $s$-particle ($s<N$)
distribution function, this irrelevant term is conventionally disregarded
using, e.g., the RPA (random phase approximation).

In general, the approximations allowing to obtain closed evolution equations
for marginals (reduced distribution functions), such as "molecular chaos", the
Bogoliubov principle of weakening of initial correlations \cite{Bogoliubov
(1946)} or RPA , actually imply selecting the specific and not very realistic
\cite{Van Kampen} uncorrelated (factorized) distribution function as an
initial state of the system, and then the problem of the "propagation of
chaos" with time should be resolved.

Thus, we see that in both approaches, a proper treating of initial
correlations is crucial for obtaining the completely closed (homogeneous)
equation for $s$-particle distribution function. It is also important for
elucidating the emergence of the irreversibility from reversible particles
dynamics. As to our understanding, there is no satisfactory arguments for
disregarding the irrelevant initial condition term \cite{Wallace}.

In order to obtain a completely closed equation for $s$-particle distribution
function one can try to include initial correlations into consideration by
transfer the initial condition term into the kernel governing the evolution in
time of this distribution function and thus to arrive at a closed kinetic
equation. This is the idea of a "subdynamics", put forward by Prigogine with
collaborators (see \cite{Balescu}), that the dynamics of $N$-particle
distribution function in the full phase space can be projected by some
projection operator on the dynamics in the subspace of selected group of $s$
particles, but no specific projection operator was suggested.

It is necessary to mention, that the authors of the work \cite{Gilbert} have
recently claimed the proving of the theorem that starting from a Newtonian
hard-sphere particle system and in the Boltzmann-Grad limit the Boltzmann
equation can be derived for the limiting case of vanishing particle density.
It is important to stress, that they claim the validity of "molecular chaos "
propagation (for $t\geq t_{0}$) in this limiting case.

In this paper, we suggest the different approach to resolving the problem,
based on the projection operator formalism. The idea is, instead of focusing
on proving the vanishing of correlations at all $t$ ("molecular chaos"
propagation) in the Boltzmann-Grad limit \cite{Gilbert}), to include them into
consideration. We show that there is a time-independent projection operator
which selects the $s$-particle ($s\leq N$) reduced distribution function
$F_{s}(t)$ and leaves unchanged an arbitrary initial distribution function
$F_{N}(0)$ of a multiparticle system of interacting classical particles. By
application of this operator to the Liouville equation for $F_{N}(t)$, we
obtain the exact completely closed (homogeneous) linear GME for a relevant
part of the $N$-particle distribution function accounting for correlations and
thereby the completely closed GME for $F_{s}(t)$. This equation differs from
the Nakajima-Zwanzig equation by absence of inhomogeneous initial correlation
term (a source), which now is "hidden" (via the introduced projection
operator) in the kernel governing the evolution of $F_{s}(t)$ .

The equation for $F_{s}(t)$ then specialized for a two-particle interaction
Hamilton function in the linear in the particles' density $n$ approximation.
As an example, the exact (in the linear in $n$ approximation) equations for a
one-particle $F_{1}(t)$ and two-particle $F_{2}(t)$ distribution functions are
obtained. They show explicitly the influence of initial correlations on the
evolution process at an arbitrary timescale.

The equation for $F_{1}(t)$ is considered in detail in the spatial homogeneous
case. It is remarkable, that on the large timescale $t\backsim t_{rel}\gg
t_{cor}$, initial correlations cease influencing the evolution of $F_{1}(t)$
and it is now described by the linear Boltzmann equation obtained with no
"molecular chaos"-type approximation. Then, it is shown, that in the time
interval
\begin{equation}
t_{cor}\ll t\ll t_{rel} \label{3}%
\end{equation}
the linear Boltzmann equation can be rewritten as the nonlinear Boltzmann
equation. It follows from (\ref{3}) that at
\begin{equation}
t_{rel}\rightarrow\infty,(l\rightarrow\infty), \label{3a}%
\end{equation}
where $l$ is the particle mean free path, the obtained nonlinear Boltzmann
equation holds for all finite times $t\gg t_{cor}$.

\section{Projection formalism}

Let us consider the Liouville equation for a distribution function
$F_{N}(t,t_{0})$ of $N$ interacting classical particles%
\begin{align}
\frac{\partial}{\partial t}F_{N}(t,t_{0})  &  =L(t)F_{N}(t,t_{0}),\nonumber\\
\int...\int dx^{N}F_{N}(t,t_{0})  &  =1,dx^{N}=dx_{1}...dx_{N}, \label{4}%
\end{align}
where $F_{N}(t,t_{0})=F_{N}(x_{1},...,x_{N};t,t_{0})$ is a function of $N$
variables $x_{i}=(\mathbf{r}_{i},\mathbf{p}_{i})$ ($i=1,...,N$) representing
the coordinates and momenta of the particles, and $L(t)$ is the Liouville
operator acting on $F_{N}(t,t_{0})$ as%

\begin{equation}
L(t)F_{N}(t,t_{0})=\{H(t),F_{N}(t,t_{0})\}_{P}=\sum_{i=1}^{N}\{\frac
{H(t)}{\partial\mathbf{r}_{i}}\frac{\partial F_{N}(t,t_{0})}{\partial
\mathbf{p}_{i}}-\frac{\partial H(t)}{\partial\mathbf{p}_{i}}\frac{\partial
F_{N}(t,t_{0})}{\partial\mathbf{r}_{i}}\}. \label{5}%
\end{equation}
Here, $\{H(t),F_{N}(t,t_{0})\}_{P}$ is the Poisson bracket and $H(t)$ is the
Hamilton function for the system under consideration generally dependent on time.

The formal solution to Eq. (\ref{4}) is
\begin{equation}
F_{N}(t,t_{0})=U(t,t_{0})F_{N}(t_{0},t_{0}), \label{6}%
\end{equation}
where the evolution operator $U(t,t_{0})$ is defined as
\begin{align}
U(t,t_{0})  &  =\exp[\int\limits_{t_{0}}^{t}d\xi L(\xi)],\nonumber\\
U(t_{0},t_{0})  &  =1. \label{7}%
\end{align}

It is practically impossible to solve Eq. (\ref{4}) for a many-particle
system. Fortunately, however, in order to calculate the measurable values
(statistical expectations) of interest, one usually only needs to know the
reduced distribution functions (marginals) $F_{s}(t,t_{0})=F_{s}%
(x_{1},...,x_{s};t,t_{0})$ dependent on much smaller number of variables
$s<<N$. In general, the $s$-particle ($s\leq N$) distribution function is
defined as \cite{Bogoliubov (1946)}
\begin{equation}
F_{s}(t,t_{0})=V^{s}\int\cdots\int dx^{\Sigma}F_{N}(t,t_{0}),dx^{\Sigma
}=dx_{s+1}...dx_{N}, \label{7a}%
\end{equation}
where $V$ is the volume of a system. From (\ref{4}) we have the normalization
condition for the reduced distribution functions $F_{s}$
\begin{equation}
\int\cdots\int dx^{s}F_{s}(t,t_{0})=V^{s},dx^{s}=dx_{1}...dx_{s}. \label{7b}%
\end{equation}
As it follows from the adopted in (\ref{7a}) and (\ref{7b}) definitions for
the multiple integrations, we divide the system of $N$ particles into a
subsystem of $s$ relevant particles and an "environment" $\Sigma$ of $N-s$
irrelevant ones. Thus, an average value of a function of the dynamic variables
of the group of $s$ particle is defined by the reduced distribution function
$F_{s}$ as%
\begin{equation}
<A_{s}>_{t,t_{0}}=\int...\int dx^{N}A_{s}F_{N}(t,t_{0})=\int...\int
dx^{s}A_{s}\frac{1}{V^{s}}F_{s}(t,t_{0}). \label{7c}%
\end{equation}

In order to obtain equations for the reduced distribution functions, it is
convenient to employ the standard projection operator technique \cite{Nakajima
(1958)}, \cite{Zwanzig (1960)}, \cite{Prigogine (1962)} and to break
$F_{N}(t,t_{0})$ into the relevant $\rho_{r}(t,t_{0})$ and irrelevant
$\rho_{i}(t,t_{0})$ parts as%
\begin{align}
F_{N}(t,t_{0})  &  =\rho_{r}(t,t_{0})+\rho_{i}(t,t_{0}),\nonumber\\
\rho_{r}(t,t_{0})  &  =PF_{N}(t,t_{0}),\rho_{i}(t,t_{0})=QF_{N}(t,t_{0}%
)=F_{N}(t,t_{0})-\rho_{r}(t,t_{0}) \label{8}%
\end{align}
with the help of some projection operators $P$ and $Q=1-P$ ($P^{2}=P$,
$Q^{2}=Q$, $P+Q=1$, $PQ=QP=0$). We note, that the relevant and irrelevant
parts depend on the coordinates and momenta of all $N$ particles in contrast
to the reduced distribution functions (like $F_{s}(t,t_{0})$). The relevant
part $\rho_{r}(t,t_{0})$ is conveniently defined in such a way that it is
related to the reduced distribution function of interest $F_{s}(t,t_{0})$,
i.e., the projection operator selects a relevant part of $F_{N}(t,t_{0})$,
which contains no correlation between a subsystem (a group of $s$ particles)
and an environment (remaining $N-s$ particles), and actually describes the
evolution of $F_{s}(t,t_{0})$. Then, the irrelevant part of the distribution
function $\rho_{i}(t,t_{0})$ contains all correlations between a subsystem and
an environment.

Applying the projection operators $P$ and $Q$ to Eq. (\ref{4}), it is easy to
obtain the equations for the relevant and irrelevant parts of $F_{N}(t)$
\begin{align}
\frac{\partial}{\partial t}\rho_{r}(t)  &  =PL(t)[\rho_{r}(t)+\rho
_{i}(t)],\nonumber\\
\frac{\partial}{\partial t}\rho_{i}(t)  &  =QL(t)[\rho_{r}(t)+\rho_{i}(t)]
\label{9}%
\end{align}
(from now on we put $t_{0}=0$). Inserting the solution of the second Eq.
(\ref{9}) for $\rho_{i}(t)$ into the first Eq. (\ref{9}), we obtain (see,
e.g., \cite{Breuer}) the conventional exact time-convolution generalized
master equation (TC-GME) known as the Nakajima-Zwanzig equation for the
relevant part of the distribution function
\begin{align}
\frac{\partial}{\partial t}\rho_{r}(t)  &  =PL(t)\rho_{r}(t)+\int
\limits_{0}^{t}PL(t)U_{Q}(t,\tau)QL(\tau)\rho_{r}(\tau)d\tau\nonumber\\
&  +PL(t)U_{Q}(t,0)\rho_{i}(0),\nonumber\\
U_{Q}(t,\tau)  &  =T\exp[%
{\displaystyle\int\limits_{\tau}^{t}}
d\xi QL(\xi)]=1+%
{\textstyle\int\limits_{\tau}^{t}}
dt_{1}\left[  QL(t_{1})\right] \nonumber\\
&  +%
{\textstyle\int\limits_{\tau}^{t}}
dt_{1}%
{\textstyle\int\limits_{t_{1}}^{t}}
dt_{2}\left[  QL(t_{2})\right]  \times\left[  QL(t_{1})\right]  +\ldots.
\label{11}%
\end{align}
Here, $T$ denotes the chronological time-ordering operator which orders the
product of time-dependent operators such that their time-arguments increase
from right to left.

This equation is quite general and valid for any initial distribution function
$F_{N}(0)$. Serving as a basis for many applications, Eq. (\ref{11}),
nevertheless, contains the undesirable and in general non-negligible
inhomogeneous term (the last term in the right hand side of (\ref{11})), which
depends via $\rho_{i\text{ }}(0)$ on the same large number of variables as the
distribution function $F_{N}(0)$ and includes all initial correlations.
Therefore, Eq. (\ref{11}) does not provide for a complete reduced description
of a multiparticle system in terms of the relevant (reduced) distribution
function. Applying Bogoliubov's principle of weakening of initial correlations
(allowing to eliminate the influence of $\rho_{i}(0)$ on the large enough time
scale $t\gg t_{cor}$) or using a factorized initial condition (like
(\ref{1})),when $\rho_{i\text{ }}(0)=QF_{N}(0)=0$ (i.e., $F_{N}(0)=\rho
_{r}(0)$), one can achieve the above-mentioned goal and obtain the homogeneous
GME for $\rho_{r}(t)$, i.e. Eq. (\ref{11}) with no initial condition term.
However, obtained in such a way homogeneous GME is either approximate and
valid only on a large enough time scale (when all initial correlations vanish)
or applicable only for a rather artificial (actually unrealistic, as pointed
in \cite{Van Kampen}) initial conditions (no correlations at an initial
instant of time).

Thus, the interesting question arises: Is it possible to obtain exact and
completely closed (homogeneous) GME, i.e. the equation with no inhomogeneous
initial correlations \ term? It means that the initial correlations would be
accounted for and contained in the kernel governing the evolution of the
relevant part of the distribution function. This would provide for the
opportunity to effectively include initial correlations into consideration on
an equal footing with collisions. To some extent this program is reminiscent
of the Progogine-Balescu idea of a subdynamics in a many-particle system
\cite{Balescu}. They attempted to find the projection operator which would
divide an $N$-particle distribution function in two independent parts: kinetic
and nonkinetic ones. The vacuum part (with no correlations) of the kinetic
part would satisfy a completely closed evolution equation with no undesirable
initial correlation term.

In what follows we will show that for an arbitrary $N$-particle initial
distribution function the Liouville linear equation can be exactly transformed
into completely closed (homogeneous) linear evolution equation for a relevant
(sufficient for calculation of the measurable values in the nonequilibrium
system) part of the $N$-particle distribution function. This can be done by
introducing a special time-independent projection operator. As a result, one
arrives at the linear completely closed equation describing the evolution of a
$s$-particle complex ($s<N$), and in this sense we can talk about a
subdynamics in a phase space of an $s$-particle subsystem. We note, that the
application of a time-independent projection operator to the Liouville
equation leads to the linear generalized master equation for the relevant part
of a distribution function.

Thus, we are looking for a completely closed equation for $\rho_{r}%
(t)=PF_{N}(t)$. The standard approach results in the inhomogeneous equation
(\ref{11}) for $\rho_{r}(t)$ with the initial correlation term $\rho
_{i}(0)=QF_{N}(0)$. If it were possible to find the projection operator such
that $PF_{N}(0)=F_{N}(0)$, then we would achieve the desired goal and obtain
exact completely closed equation for $\rho_{r}(t)$.

\section{Subdynamics}

Let us introduce the following projection operator%

\begin{equation}
P=P_{s\Sigma}=\frac{F_{N}(0)}{F_{s}(0)}V^{s}\int\ldots\int dx^{\Sigma},
\label{12}%
\end{equation}
where $F_{N}(0)=F_{N}(x_{1},...,x_{N};0)$ is an initial $N$-particle
distribution function depending on all $N$ variables $x_{i}$ and $F_{s}(0)$ is
an $s$-particle distribution function at $t=0$ defined by (\ref{7a}). One can
see that
\begin{equation}
\int\ldots\int dx^{\Sigma}\frac{F_{N}(0)}{F_{s}(0)}V^{s}=1, \label{13}%
\end{equation}
which proves that $P_{s\Sigma}$ is a projector. Moreover, acting by the
projector (\ref{12}) on the initial distribution $F_{N}(0)$ we find that due
to (\ref{13})
\begin{equation}
P_{s\Sigma}F_{N}(0)=\frac{F_{N}(0)}{F_{s}(0)}V^{s}\int\ldots\int dx^{\Sigma
}F_{N}(0)=F_{N}(0). \label{14}%
\end{equation}
We see that an initial distribution $F_{N}(0)$ remains unchanged under the
action of the projector (\ref{12}). Therefore, using the projector (\ref{12})
for the derivation of an equation for the relevant part of the distribution
function
\begin{equation}
f_{r}^{s}(t)=P_{s\Sigma}F_{N}(t)=\frac{F_{N}(0)}{F_{s}(0)}V^{s}\int
dx^{\Sigma}F_{N}(t)=\frac{F_{N}(0)}{F_{s}(0)}F_{s}(t), \label{14a}%
\end{equation}
we arrive (instead of Eq. (\ref{11})) at the following exact completely closed
(homogeneous) equation%

\begin{align}
\frac{\partial}{\partial t}f_{r}^{s}(t)  &  =P_{s\Sigma}L(t)f_{r}^{s}%
(t)+\int\limits_{0}^{t}P_{s\Sigma}L(t)U_{Q_{s\Sigma}}(t,\tau)Q_{s\Sigma}%
L(\tau)f_{r}^{s}(\tau)d\tau,\nonumber\\
U_{Q_{s\Sigma}}(t,\tau)  &  =T\exp[%
{\displaystyle\int\limits_{\tau}^{t}}
d\xi Q_{s\Sigma}L(\xi)],Q_{s\Sigma}=1-P_{s\Sigma}. \label{15}%
\end{align}

Equation (\ref{15}) is the main result of this Section. It means, that if we
start with the arbitrary initial distribution function $F_{N}(0)$ and use the
projector (\ref{12}), the subdynamics in the subspace of the group of selected
$s$ particles shows up. It is especially evident if we rewrite Eq. (\ref{15})
for the reduced distribution function $F_{s}(t)$, using the definition
(\ref{14a}), i.e.,%
\begin{align}
\frac{\partial}{\partial t}\frac{1}{V^{s}}F_{s}(t)  &  =\int...\int
dx^{\Sigma}L(t)\frac{F_{N}(0)}{F_{s}(0)}F_{s}(t)\nonumber\\
&  +\int...\int dx^{\Sigma}\int\limits_{0}^{t}L(t)U_{Q_{s\Sigma}}%
(t,\tau)Q_{s\Sigma}L(\tau)\frac{F_{N}(0)}{F_{s}(0)}F_{s}(\tau)d\tau.
\label{15a}%
\end{align}

Thus, Eq. (\ref{15a}) resolves the problem of finding the completely closed
equation for the reduced distribution function $F_{s}(t)$ with no
approximation of the "molecular chaos" type or any other. An average value of
a function of the dynamic variables of the group of $s$ particle is defined by
the reduced distribution function $F_{s}(t)$ as%
\begin{equation}
<A_{s}>_{t}=\int...\int dx^{N}A_{s}F_{N}(t)=\int...\int dx^{s}A_{s}\frac
{1}{V^{s}}F_{s}(t)=\int...\int dx^{N}A_{s}f_{r}^{s}(t). \label{17}%
\end{equation}

We present the system's Hamilton function as%
\begin{equation}
H=H_{s}+H_{\Sigma}+H_{s\Sigma}, \label{17a}%
\end{equation}
\ where $H_{s}$ is the Hamilton function of the selected group of $s$
particles, which interacts (via $H_{s\Sigma}$) with an environment $\Sigma$ of
other $N-s$ particles (described by $H_{\Sigma}$). The corresponding Liouville
operator is $L=L_{s}+L_{\Sigma}+L_{s\Sigma}$.

The condition (\ref{13}) also leads to the following relations
\begin{align}
\int\ldots\int dx^{\Sigma}Q_{s\Sigma}  &  =0,\int\ldots\int dx^{\Sigma}%
f_{i}^{s}(t)=0,\nonumber\\
\int\ldots\int dx^{\Sigma}U_{Q_{s\Sigma}}(t,t_{1})Q_{s\Sigma}  &  =0,f_{i}%
^{s}(t)=Q_{s\Sigma}F_{N}(t), \label{23c}%
\end{align}
which can be used for simplifying Eqs. (\ref{15}) and (\ref{15a}).

Let us now specify the time-independent Hamilton function $H$ for the case of
the identical particles with the two-body interparticle interaction as%
\begin{align}
H  &  =H_{S}+H_{\Sigma}+H_{s\Sigma},\nonumber\\
H_{s}  &  =\sum\limits_{i=1}^{s}\frac{\mathbf{p}_{i}^{2}}{2m}+\sum
\limits_{1\leq i<j\leq s}V_{ij}(\left\vert \mathbf{r}_{i}-\mathbf{r}%
_{j}\right\vert ),\nonumber\\
H_{\Sigma}  &  =%
{\displaystyle\sum\limits_{i=s+1}^{N}}
\frac{\mathbf{p}_{i}^{2}}{2m}+\sum\limits_{s+1\leq i<j\leq N}V_{ij}(\left\vert
\mathbf{r}_{i}-\mathbf{r}_{j}\right\vert ),\nonumber\\
H_{s\Sigma}  &  =\sum\limits_{i=1}^{s}\sum\limits_{j=s+1}^{N}V_{ij}(\left\vert
\mathbf{r}_{i}-\mathbf{r}_{j}\right\vert ). \label{24}%
\end{align}
The corresponding to (\ref{24}) Liouville operator $L$ is%
\begin{align}
L  &  =L_{s}+L_{\Sigma}+L_{s\Sigma},\nonumber\\
L_{s}  &  =-%
{\displaystyle\sum\limits_{i=1}^{s}}
\mathbf{v}_{i}\mathbf{\nabla}_{i}+\sum\limits_{1\leq i<j\leq s}(\mathbf{\nabla
}_{i}V_{ij})\cdot(\frac{\partial}{\partial\mathbf{p}_{i}}-\frac{\partial
}{\partial\mathbf{p}_{j}}),\nonumber\\
L_{\Sigma}  &  =-%
{\displaystyle\sum\limits_{i=s+1}^{N}}
\mathbf{v}_{i}\mathbf{\nabla}_{i}+\sum\limits_{s+1\leq i<j\leq N}%
(\mathbf{\nabla}_{i}V_{ij})\cdot(\frac{\partial}{\partial\mathbf{p}_{i}}%
-\frac{\partial}{\partial\mathbf{p}_{j}}),\nonumber\\
L_{s\Sigma}  &  =\sum\limits_{i=1}^{s}\sum\limits_{j=s+1}^{N}(\mathbf{\nabla
}_{i}V_{ij})\cdot(\frac{\partial}{\partial\mathbf{p}_{i}}-\frac{\partial
}{\partial\mathbf{p}_{j}}),\nonumber\\
\mathbf{v}_{i}  &  =\mathbf{p}_{i}/m,\mathbf{\nabla}_{i}=\frac{\partial
}{\partial\mathbf{r}_{i}},V_{ij}=V_{ij}(\left\vert \mathbf{r}_{i}%
-\mathbf{r}_{j}\right\vert ). \label{25}%
\end{align}

As usual, we also assume that all functions $\Phi(x_{1},\ldots,x_{N})$ defined
on the phase space and their derivatives vanish at the boundaries of the
configuration space and at $\mathbf{p}_{i}=\pm\infty$. These boundary
conditions and the explicit form of the Liouville operators (\ref{25}) lead to
the following relations
\begin{align}
\int...\int dx^{\Sigma}L_{\Sigma}\Phi(x_{1},\ldots,x_{N})  &  =0,\nonumber\\
\int\ldots\int dx^{\Sigma}L_{s\Sigma}\Phi(x_{1},\ldots,x_{N})  &  =\int
\ldots\int dx^{\Sigma}\sum\limits_{i=1}^{s}\sum\limits_{j=s+1}^{N}%
(\mathbf{\nabla}_{i}V_{ij})\cdot\frac{\partial}{\partial\mathbf{p}_{i}}%
\Phi(x_{1},\ldots,x_{N}). \label{25a}%
\end{align}

Using (\ref{12}), (\ref{13}) and (\ref{25a}), one can prove the following
relation%
\begin{equation}
U_{Q_{s\Sigma}}(t,t_{1})Q_{s\Sigma}=\overline{U}_{Q_{s\Sigma}}(t,t_{1}%
)Q_{s\Sigma}, \label{26b}%
\end{equation}
where
\begin{align}
\overline{U}_{Q_{s\Sigma}}(t,t_{1})  &  =\exp[(L_{0}+Q_{s\Sigma}L_{s\Sigma
})(t-t_{1})],\nonumber\\
L_{0}  &  =L_{s}+L_{\Sigma}. \label{27}%
\end{align}

Making use of (\ref{13}), (\ref{23c}), (\ref{25a}), and (\ref{26b} ), Eq.
(\ref{15a}) can be rewritten as the following equation for the $s$-particle
distribution function of interest $F_{s}(t)$%
\begin{align}
\frac{\partial}{\partial t}F_{s}(t)  &  =[L_{s}+V^{s}\int\ldots\int
dx^{\Sigma}L_{s\Sigma}\frac{F_{N}(0)}{F_{s}(0)}]F_{s}(t)\nonumber\\
&  +V^{s}\int\ldots\int dx^{\Sigma}\int\limits_{0}^{t}L_{s\Sigma}\overline
{U}_{Q_{s\Sigma}}(t,\tau)\left\{  [Q_{s\Sigma}L_{s\Sigma}+L_{\Sigma}%
]\frac{F_{N}(0)}{F_{s}(0)}+[L_{s},\frac{F_{N}(0)}{F_{s}(0)}]\right\}
F_{s}(\tau)d\tau, \label{28}%
\end{align}
where $[A,B]$ is the commutator of $A$ and $B$.

Simplified Eq. (\ref{28}) is the main result of this section. It is the linear
time-convolution homogeneous GME obtained from the linear Liouville equation
for $N$-particle distribution function $F_{N}(x_{1},x_{2},\ldots,x_{N};t)$ by
means of the introduced projection operator (\ref{12}). Influence of initial
correlations on all terms in Eq. (\ref{28}) is given by the factor
$\frac{F_{N}(0)}{F_{s}(0)}$ which resulted from the action of the projection
operator $P_{s\Sigma}$. The first term $L_{s}F_{s}(t)$ is a conventional flow
term. The second term $V^{s}\int dx^{\Sigma}L_{s\Sigma}\frac{F_{N}(0)}%
{F_{s}(0)}F_{s}(t)$ represents the (reminiscent of Vlasov's) field acting on
the subsystem of $s$ particles and determined by the environment of $N-s$
particles. The third term is a collision term with additional factor
$L_{\Sigma}\frac{F_{N}(0)}{F_{s}(0)}$ accounting for the evolution of the
irrelevant subsystem of $N-s$ particles, and the last term exists only due to
initial correlations. In the absence of initial correlations, when $F_{N}(0)$
is given by%

\begin{equation}
F_{N}(0)=F_{s}(0)F_{N-s}(0),F_{N-s}(0)=\rho_{\Sigma}(0), \label{28a}%
\end{equation}
where $\rho_{\Sigma}(0)$ is the initial distribution function for an
environment of $N-s$ particles, this term vanishes.

\section{Equation for s-particle distribution function in linear in particles'
density approximation}

Let us present the terms of Eq. (\ref{28}) in a more explicit form. Using
(\ref{7a}), (\ref{25}), and the symmetry of the distribution functions with
regards to the variables $x_{i}$, the "Vlasov" term can be exactly rewritten
as
\begin{align}
V^{s}\int\ldots\int dx^{\Sigma}L_{s\Sigma}\frac{F_{N}(0)}{F_{s}(0)}F_{s}(t)
&  =V^{s}\int\ldots\int dx^{\Sigma}\sum\limits_{i=1}^{s}\sum\limits_{j=s+1}%
^{N}L_{ij}\frac{F_{N}(0)}{F_{s}(0)}F_{s}(t)\nonumber\\
&  =V^{s}(N-s)\sum\limits_{i=1}^{s}\int dx_{s+1}\ldots\int dx_{N}%
L_{i,s+1}\frac{F_{N}(x_{1},x_{2},\ldots,x_{N};0)}{F_{s}(x_{1},x_{2}%
,\ldots,x_{s};0)}F_{s}(t)\nonumber\\
&  =\frac{N-s}{V}\sum\limits_{i=1}^{s}\int dx_{s+1}L_{i,s+1}\frac{F_{s+1}%
(0)}{F_{s}(0)}F_{s}(t),\nonumber\\
L_{ij}  &  =(\mathbf{\nabla}_{i}V_{ij})\cdot(\frac{\partial}{\partial
\mathbf{p}_{i}}-\frac{\partial}{\partial\mathbf{p}_{j}}). \label{29}%
\end{align}
We remind that $F_{s}(t)=F_{s}(x_{1},x_{2},\ldots,x_{s};t)$. At $N\gg s$, this
term is proportional to the particles' density $n=N/V$. Derivation of
(\ref{29}) suggests that the forms of the Liouville operators (\ref{25}) and
integral operator $\int\ldots\int dx^{\Sigma}$ provide an opportunity for
expanding the other terms of Eq. (\ref{28}) into the $N-s$ "environment"
particles' density $(N-s)/V$ series. One can easily see, that application of
the integral operator $\int\ldots\int dx^{\Sigma}$ (or $P_{s\Sigma}$) leads to
the expressions of at least first order in $(N-s)/V$.

Let us consider equation (\ref{28}) in the first approximation in the $N-s$
particles' density $(N-s)/V$. The corresponding dimensionless small parameter
of a perturbation expansion is
\begin{equation}
\gamma=r_{0}^{3}(N-s)/V\ll1 \label{30}%
\end{equation}
where $r_{0}$ is the effective radius of an inter-particle interaction. The
condition (\ref{30}) allows for introducing the different time scales: the
correlation time $t\thicksim$ $t_{cor}\thicksim r_{0}/v$ ($v$ is the
characteristic particle velocity) and the relaxation time $t\thicksim t_{rel}$
for a one particle distribution function with $t_{rel}\gg t_{cor}$. In the
first in $\gamma$ approximation, all terms in Eq. (\ref{28}) with $P_{s\Sigma
}$ can be neglected (in particular, in $\overline{U}(t,t^{^{\prime}})$)
because there is already a common prefactor $\int\ldots\int dx^{\Sigma
}L_{s\Sigma}$. In this approximation the operator $\overline{U}(t,t^{^{\prime
}})$ can be reduced to
\begin{equation}
\overline{U}(t,t^{\prime})=\exp[(L_{s}+L_{\Sigma}+L_{s\Sigma})(t-t^{\prime})].
\label{31}%
\end{equation}
As the result, we get the following expression for the collision term of Eq.
(\ref{28}) in the linear approximation in the particles' density%
\begin{align}
&  V^{s}\int\ldots\int dx^{\Sigma}\int\limits_{0}^{t}L_{s\Sigma}\overline
{U}_{Q_{s\Sigma}}(t,\tau)\left\{  [Q_{s\Sigma}L_{s\Sigma}+L_{\Sigma}%
]\frac{F_{N}(0)}{F_{s}(0)}+[L_{s},\frac{F_{N}(0)}{F_{s}(0)}]\right\}
F_{s}(\tau)d\tau\nonumber\\
&  =\frac{N-s}{V}\sum\limits_{i=1}^{s}\int dx_{s+1}L_{i,s+1}%
{\textstyle\int\limits_{0}^{t}}
d\tau\exp[L_{s+1}(t-\tau)]\{(\sum\limits_{j=1}^{s}L_{j,s+1}+L_{s+1}^{0}%
)\frac{F_{s+1}(0)}{F_{s}(0)}+[L_{s},\frac{F_{s+1}(0)}{F_{s}(0)}]\}F_{s}%
(\tau),\nonumber\\
L_{i}^{0}  &  =-\mathbf{v}_{i}\mathbf{\nabla}_{i}. \label{32}%
\end{align}

Thus, Eqs. (\ref{28}), (\ref{29}) and (\ref{32}) define the evolution equation
for the $s$-particle distribution function $F_{s}(x_{1},...,x_{s};t)$ in the
linear in $(N-s)/V$ approximation. This equation is completely closed in a
sense, that it does not contain a source (like that in Eq. (\ref{11}))
accounting for initial (at $t=t_{0}$) correlations. It is worth noting that at
$s=N$, the terms defined by (\ref{29}) and (\ref{32}) vanish and Eq.
(\ref{28}) reduces to the Liouville evolution for $N$-particle distribution
function%
\begin{equation}
\frac{\partial F_{N}(t)}{\partial t}=L_{N}F_{N}(t), \label{33}%
\end{equation}
as one would expect.

Note, that the solution of Eq. (\ref{28}) in the first in $n$ approximation to
the kernel (given by (\ref{29}) and (\ref{32})) generally contains an infinite
subseries of terms of any power of $n$ (this is an advantage of the expanding
of an equation's kernel as compared to the expansion of the solution itself).

To discuss the obtained result in more detail, let us write down explicitly
the equations for one-particle and two-particle distribution functions, which
are mostly needed for applications. They are as follows%

\begin{align}
\frac{\partial F_{1}(x_{1};t)}{\partial t}  &  =L_{1}^{0}F_{1}(x_{1};t)+n\int
dx_{2}L_{12}\frac{F_{2}(x_{1},x_{2};0)}{F_{1}(x_{1};0)}F_{1}(x_{1}%
;t)\nonumber\\
&  +n\int dx_{2}L_{12}%
{\textstyle\int\limits_{0}^{t}}
d\tau e^{L_{2}(t-\tau)}\{(L_{12}+L_{2}^{0})\frac{F_{2}(x_{1},x_{2};0)}%
{F_{1}(x_{1};0)}\nonumber\\
&  +[L_{1}^{0},\frac{F_{2}(x_{1},x_{2};0)}{F_{1}(x_{1};0)}]\}F_{1}(x_{1}%
;\tau), \label{34}%
\end{align}
and%

\begin{align}
\frac{\partial F_{2}(x_{1},x_{2};t)}{\partial t}  &  =L_{2}F_{2}(x_{1}%
,x_{2};t)+n\int dx_{3}(L_{13}+L_{23})\frac{F_{3}(x_{1},x_{2},x_{3};0)}%
{F_{2}(x_{1},x_{2};0)}F_{2}(x_{1},x_{2};t)\nonumber\\
&  +n\int dx_{3}(L_{13}+L_{23})%
{\textstyle\int\limits_{0}^{t}}
d\tau e^{L_{3}(t-\tau)}[(L_{13}+L_{23}+L_{3}^{0})\frac{F_{3}(x_{1},x_{2}%
,x_{3};0)}{F_{2}(x_{1},x_{2};0)}\nonumber\\
&  +[L_{2},\frac{F_{3}(x_{1},x_{2},x_{3};0)}{F_{2}(x_{1},x_{2};0)}%
]\}F_{2}(x_{1},x_{2};\tau), \label{35}%
\end{align}

Equations (\ref{34}) and (\ref{35}) are exact in the linear approximation in
the small density parameter (\ref{30}) and is valid on any timescale and for
any space-inhomogeneity of the system under consideration. The evolution in
time is governed either by the exact two-particle propagator $G_{2}%
(t)=e^{L_{2}t}=\exp[(L_{1}^{0}+L_{2}^{0}+L_{12})t]$ in Eq. (\ref{34}) or by
the exact three-particle propagator $G_{3}(t)=e^{L_{3}t}=\exp[(L_{1}^{0}%
+L_{2}^{0}+L_{3}^{0}+L_{12}+L_{13}+L_{23})t]$ in Eq. (\ref{35}) (it is natural
for the considered dilute gas in the lowest approximation in the density).

We note, that Eqs. (\ref{34}), (\ref{35}) slightly remind the BBGKY hierarchy
by appearance of the distribution functions, $F_{2}(x_{1},x_{2};0)$,
$F_{3}(x_{1},x_{2},x_{3};0)$, in the closed equations for $F_{1}(x_{1};t)$ and
$F_{2}(x_{1},x_{2};t)$ (in the kernels of these equations), respectively, but
this hierarchy relates only to the initial values of the distribution functions.

Let us consider Eq. (\ref{34}) for $F_{1}(x_{1};t)$ in more detail. In order
to reveal the explicit contribution of initial correlations and make these
equations look more familiar, it is useful to present the initial
many-particle distribution functions in terms of the vacuum (not correlated)
and irreducible correlations forms. Thus, let us rewrite the initial
correlation factor in Eq. (\ref{34}) as%
\begin{align}
\frac{F_{2}(x_{1},x_{2};0)}{F_{1}(x_{1};0)}  &  =\frac{F_{1}(x_{1}%
;0)F_{1}(x_{2};0)+g_{2}(x_{1},x_{2};0)}{F_{1}(x_{1};0)}\nonumber\\
&  =F_{1}(x_{2};0)[1+\frac{g_{2}(x_{1},x_{2};0)}{F_{1}(x_{1};0)F_{1}(x_{2}%
;0)}], \label{36}%
\end{align}
where $g_{2}(x_{1},x_{2};0)$ is an irreducible two-particle correlation
function. Then, Eq. (\ref{34}) can be presented as the following linear
equation for one-particle distribution function $F_{1}(x_{1};t)$%
\begin{align}
\frac{\partial F_{1}(x_{1};t)}{\partial t}  &  =L_{1}^{0}F_{1}(x_{1};t)+n\int
dx_{2}L_{12}[1+C_{2}(x_{1},x_{2};0)]F_{1}(x_{2};0)F_{1}(x_{1};t)\nonumber\\
&  +n\int dx_{2}L_{12}%
{\textstyle\int\limits_{0}^{t}}
dt_{1}e^{L_{2}t_{1}}\{L_{12}[1+C_{2}(x_{1},x_{2};0)]+L_{2}^{0}\nonumber\\
&  +(L_{1}^{0}+L_{2}^{0})C_{2}(x_{1},x_{2};0)-C_{2}(x_{1},x_{2};0)L_{1}%
^{0}\}F_{1}(x_{2};0)F_{1}(x_{1};t-t_{1}), \label{37}%
\end{align}
where we introduced the dimensionless parameter of initial correlations
\begin{equation}
C_{2}(x_{1},x_{2};0)=\frac{g_{2}(x_{1},x_{2};0)}{F_{1}(x_{1};0)F_{1}(x_{2};0)}
\label{38}%
\end{equation}
and changed the time integration variable $t-\tau\rightarrow t_{1}$.

Let us consider a spatially homogeneous case. In this case a one-particle
distribution function does not depend on a particle coordinate, $F_{1}%
(x_{j},t)=F_{1}(\mathbf{p}_{j},t)$ ($\int F_{1}(\mathbf{p,}t)d\mathbf{p=}1$),
and the multiparticle correlation functions depend only on the differences of
coordinates, e.g., $g_{2}(x_{i},x_{j};0)=g_{2}(\mathbf{r}_{i}-\mathbf{r}%
_{j},\mathbf{p}_{1},\mathbf{p}_{2};0)$. Remembering the definition of
$L_{i}^{0}$, and that a potential $V_{ij}$ depends on the particle coordinates
difference (see (\ref{25})), we obtain from Eq. (\ref{37}) the following
equation%
\begin{align}
\frac{\partial F_{1}(\mathbf{p}_{1};t)}{\partial t}  &  =n\int d\mathbf{r}%
_{2}\int d\mathbf{p}_{2}L_{12}C_{2}(\mathbf{r}_{1}-\mathbf{r}_{2}%
,\mathbf{p}_{1},\mathbf{p}_{2};0)F_{1}(\mathbf{p}_{2};0)F_{1}(\mathbf{p}%
_{1};t)\nonumber\\
&  +n\int d\mathbf{r}_{2}\int d\mathbf{p}_{2}L_{12}%
{\textstyle\int\limits_{0}^{t}}
dt_{1}e^{L_{2}t_{1}}\{L_{12}[1+C_{2}(\mathbf{r}_{1}-\mathbf{r}_{2}%
,\mathbf{p}_{1},\mathbf{p}_{2};0)]\nonumber\\
&  -(\mathbf{v}_{1}-\mathbf{v}_{2})\mathbf{\nabla}_{1}C_{2}(\mathbf{r}%
_{1}-\mathbf{r}_{2},\mathbf{p}_{1},\mathbf{p}_{2};0)\}F_{1}(\mathbf{p}%
_{2};0)]F_{1}(\mathbf{p}_{1};t-t_{1}),\nonumber\\
C_{2}(\mathbf{r}_{1}-\mathbf{r}_{2},\mathbf{p}_{1},\mathbf{p}_{2};0)  &
=\frac{g_{2}(\mathbf{r}_{1}-\mathbf{r}_{2},\mathbf{p}_{1},\mathbf{p}_{2}%
;0)}{F_{1}(\mathbf{p}_{1};0)F_{1}(\mathbf{p}_{2};0)}, \label{39}%
\end{align}
where we have used the definition $L_{i}^{0}=-\mathbf{v}_{i}\mathbf{\nabla
}_{i}$ and that $\frac{\partial}{\partial\mathbf{r}_{1}}g_{2}(\mathbf{r}%
_{1}-\mathbf{r}_{2},\mathbf{p}_{1},\mathbf{p}_{2};0)=-\frac{\partial}%
{\partial\mathbf{r}_{2}}g_{2}(\mathbf{r}_{1}-\mathbf{r}_{2},\mathbf{p}%
_{1},\mathbf{p}_{2};0)$.

\section{Large timescale equation}

Equation (\ref{39}) is valid on all timescales. We will be interested in
considering the large timescale $t\sim t_{rel}$ on which the one-particle
distribution function $F_{1}(x_{1};t)$ changes. Due to the low density
condition (\ref{30}), there is the time hierarchy%
\begin{equation}
t_{cor}\ll t_{rel}, \label{40}%
\end{equation}
where $t_{cor}$ is the particles correlation time and $t_{rel}\thicksim
\gamma^{-1}t_{cor}$.

\bigskip The two-particle propagator $U_{2}(t)=e^{L_{2}t}$ in Eq. (\ref{39})
satisfies the integral equation%

\begin{equation}
U_{2}(t)=U_{2}^{0}(t)+\int\limits_{0}^{t}dt_{1}U_{2}^{0}(t-t_{1})L_{12}%
U_{2}(t_{1}),\text{ } \label{41}%
\end{equation}
where $U_{2}^{0}(t)=\exp(L_{1}^{0}+L_{2}^{0})t$ is the propagator for
noninteracting particles. The action of $N$-particle "free" propagator
$U_{N}^{0}(t)$ on any function defined on the phase space is given by%
\begin{align}
e^{L_{N}^{0}t}\Phi(x_{1},\ldots,x_{N},t)  &  =\left[  \prod\limits_{i=1}%
^{N}e^{L_{i}^{0}t}\right]  \Phi(x_{1},\ldots,x_{N},t)\nonumber\\
&  =\Phi(\mathbf{x}_{1}-\mathbf{v}_{1}t,\mathbf{p}_{1},\ldots,\mathbf{x}%
_{N}-\mathbf{v}_{N}t,\mathbf{p}_{N},t), \label{42}%
\end{align}
Hence, one can expect that under the action of the two-particle propagator in
the integral of Eq. (\ref{39}) the distance between particles $\left\vert
\mathbf{r}_{i}-\mathbf{r}_{j}\right\vert $ in the two-particle interaction
$V_{ij}(\left\vert \mathbf{r}_{i}-\mathbf{r}_{j}\right\vert )$ increases (with
overwhelming probability) with time and $V_{ij}(\left\vert \mathbf{r}%
_{i}-\mathbf{r}_{j}\right\vert )$ vanishes at a distance $\left\vert
\mathbf{r}_{i}-\mathbf{r}_{j}\right\vert >r_{0}$ at the characteristic time
$t_{cor}\backsim r_{0}/v$, where $v$ is a mean particle velocity. Thus, if the
effective particle interaction $V_{ij}(\left\vert \mathbf{r}_{i}%
-\mathbf{r}_{j}\right\vert )$ has a finite range and vanishes at a distance
$\left\vert \mathbf{r}_{i}-\mathbf{r}_{j}\right\vert >r_{0}$, the integrand of
the integral over $t_{1}$ in Eq. (\ref{39}) vanishes for $t>t_{cor}$.

Therefore, it is reasonable to assume that the memory kernel in (\ref{39})
vanishes rapidly on the kinetic timescale $t\gtrsim t_{rel}\gg t_{cor}$ due to
the damping of correlations resulting from collisions (the Markovian
approximation). Accordingly, the upper limit of integration over $t_{1}$ in
(\ref{39}) can be extended to infinity and $F_{1}(x_{1};t-t_{1})$ can be
replaced with $F_{1}(x_{1};t)$ (the integration over $t_{1}$ in (\ref{39})
gives essential contribution only up to $t_{1}\thicksim t_{cor}\ll t\thicksim
t_{rel}$). Thus, Eq. (\ref{39}) can be approximated by the following
time-local equation%
\begin{align}
\frac{\partial F_{1}(\mathbf{p}_{1};t)}{\partial t}  &  =n\int d\mathbf{r}%
_{2}\int d\mathbf{p}_{2}L_{12}C_{2}(\mathbf{r}_{1}-\mathbf{r}_{2}%
,\mathbf{p}_{1},\mathbf{p}_{2};0)F_{1}(\mathbf{p}_{2};0)F_{1}(\mathbf{p}%
_{1};t)\nonumber\\
&  =n\int d\mathbf{r}_{2}\int d\mathbf{p}_{2}L_{12}%
{\textstyle\int\limits_{0}^{\infty}}
dt_{1}U_{2}(t_{1})\{L_{12}[1+C_{2}(\mathbf{r}_{1}-\mathbf{r}_{2}%
,\mathbf{p}_{1},\mathbf{p}_{2};0)]F_{1}(\mathbf{p}_{2};0)F_{1}(\mathbf{p}%
_{1};t)\nonumber\\
&  -(\mathbf{v}_{1}-\mathbf{v}_{2})\mathbf{\nabla}_{1}C_{2}(\mathbf{r}%
_{1}-\mathbf{r}_{2},\mathbf{p}_{1},\mathbf{p}_{2};0)\}F_{1}(\mathbf{p}%
_{2};0)]F_{1}(\mathbf{p}_{1};t),\nonumber\\
t  &  \gg t_{cor}. \label{43}%
\end{align}

On the basis of the same arguments, we can also expect that the influence of
initial correlations on the evolution of $F_{1}(\mathbf{p}_{1};t)$, given by
the correlation function $\frac{g_{2}(\mathbf{x}_{1}-\mathbf{x}_{2}%
,\mathbf{p}_{i},\mathbf{p}_{j};0)}{F_{1}(\mathbf{p}_{1};0)F_{1}(\mathbf{p}%
_{2};0)}$ in the second and third lines of Eq. (\ref{43}), is effective in the
time interval $0\leq t_{1}\leq t_{cor}$ for the finite interparticle
interaction range.

\section{Influence of initial correlations}

Let us consider the terms of Eq. (\ref{43}) related to initial correlations.
It is convenient to use the variables $\mathbf{v}_{i}=\mathbf{p}_{i}/m$,
$\mathbf{r}=\mathbf{r}_{1}-\mathbf{r}_{2}$, and $\mathbf{g}=\mathbf{v}%
_{1}-\mathbf{v}_{2}$. Then, the first term in the r.h.s. of (\ref{43}) can be
presented as \
\begin{align}
&  n\int d\mathbf{r}\int d\mathbf{v}_{2}L^{\prime}\varphi_{c}(\mathbf{v}%
_{1},\mathbf{v}_{2},\mathbf{r};t),\nonumber\\
\varphi_{c}(\mathbf{v}_{1},\mathbf{v}_{2},\mathbf{r};t)  &  =C_{2}%
(\mathbf{r},\mathbf{v}_{1},\mathbf{r}_{2};0)F_{1}(\mathbf{v}_{2}%
;0)F_{1}(\mathbf{v}_{1};t), \label{43'}%
\end{align}
where
\begin{align}
L^{\prime}  &  =[\mathbf{\nabla}V(\mathbf{r})]\cdot\mathbf{\partial,\nabla
=}\frac{\partial}{\partial\mathbf{r}},\partial=\frac{2}{m}\frac{\partial
}{\partial\mathbf{g}},\mathbf{g=v}_{1}-\mathbf{v}_{2},\nonumber\\
C_{2}(\mathbf{r},\mathbf{v}_{1},\mathbf{v}_{2};0)  &  =\frac{g_{2}%
(\mathbf{r},\mathbf{v}_{1},\mathbf{v}_{2};0)}{F_{1}(\mathbf{v}_{1}%
;0)F_{1}(\mathbf{v}_{2};0)},\nonumber
\end{align}
and we have used that%
\begin{equation}
L_{12}=(\mathbf{\nabla}_{1}V_{12})\cdot(\frac{\partial}{\partial\mathbf{p}%
_{1}}-\frac{\partial}{\partial\mathbf{p}_{2}}). \label{43'''}%
\end{equation}

The correlation term in the second line of Eq. (\ref{43}) contributing to the
collision integral can be rewritten as%

\begin{align}
&  n\int d\mathbf{v}_{2}\int J_{c}(\mathbf{v}_{1},\mathbf{v}_{2}%
;t),\nonumber\\
J_{c}(\mathbf{v}_{1},\mathbf{v}_{2};t)  &  =\int d\mathbf{r}L^{\prime}%
{\textstyle\int\limits_{0}^{\infty}}
dt_{1}U_{2}(t_{1})L^{\prime}\varphi_{c}(\mathbf{v}_{1},\mathbf{v}%
_{2},\mathbf{r};t), \label{43''''}%
\end{align}
and the last correlation term in Eq. (\ref{43}) can be written as%
\begin{equation}
n\int d\mathbf{r}\int d\mathbf{v}_{2}%
{\textstyle\int\limits_{0}^{\infty}}
dt_{1}L^{\prime}U_{2}(t_{1})[-\mathbf{g\bullet\nabla]}\varphi_{c}%
(\mathbf{v}_{1},\mathbf{v}_{2},\mathbf{r};t). \label{44}%
\end{equation}

Following the approach of \cite{Balescu}, let us consider the terms
(\ref{43''''}) and (\ref{44}) conditioned by initial correlations and
containing the time propagator $U_{2}(t_{1})$. It can be shown \cite{Balescu},
that the integral over time $%
{\textstyle\int\limits_{0}^{\infty}}
dt_{1}U_{2}(t_{1})$ can be presented as (see (\ref{41}))
\begin{equation}
Z=G+GL^{\prime}Z,G=\lim_{p\rightarrow+0}%
{\displaystyle\int\limits_{0}^{\infty}}
d\tau e^{-p\tau}U_{2}^{0}(\tau),Z=\lim_{p\rightarrow+0}%
{\displaystyle\int\limits_{0}^{\infty}}
d\tau e^{-p\tau}U_{2}(\tau). \label{46}%
\end{equation}
In the matrix form, $Z(\mathbf{r},\mathbf{g};\mathbf{r}^{\prime}%
,\mathbf{g}^{\prime})$ satisfies the equation
\begin{equation}
\{\mathbf{g\cdot\nabla-}[\mathbf{\nabla}V(\mathbf{r})]\cdot\mathbf{\partial
}\}Z(\mathbf{r},\mathbf{g};\mathbf{r}^{\prime},\mathbf{g}^{\prime}%
)=\delta(\mathbf{r}-\mathbf{r}^{\prime})\delta(\mathbf{g}-\mathbf{g}^{\prime
}), \label{47}%
\end{equation}
i.e., it is Green's function of the two-particle Liouville operator (see
(\ref{25})), whereas the matrix $G(\mathbf{r},\mathbf{g};\mathbf{r}^{\prime
},\mathbf{g}^{\prime})$ (see (\ref{42})) is diagonal with respect to velocity
indexes $G(\mathbf{r},\mathbf{g};\mathbf{r}^{\prime},\mathbf{g}^{\prime
})=G^{0}(\mathbf{r}-\mathbf{r}^{\prime})\delta(\mathbf{g}-\mathbf{g}^{\prime
})$ and $G^{0}(\mathbf{r}-\mathbf{r}^{\prime})$ is Green's function of the
unperturbed Liouville equation%
\begin{equation}
\mathbf{g\cdot\nabla}G^{0}(\mathbf{r}-\mathbf{r}^{\prime})=\delta
(\mathbf{r}-\mathbf{r}^{\prime}). \label{48}%
\end{equation}
If we introduce the function%
\begin{equation}
f_{c}(\mathbf{r},\mathbf{g;}t)=\varphi_{c}(\mathbf{v}_{1},\mathbf{v}%
_{2},\mathbf{r};t)+\int d\mathbf{r}^{\prime}\int d\mathbf{g}^{\prime
}Z(\mathbf{r},\mathbf{g};\mathbf{r}^{\prime},\mathbf{g}^{\prime}%
)[\mathbf{\nabla}^{\prime}V(\mathbf{r}^{\prime})]\cdot\mathbf{\partial
}^{\prime}\varphi_{c}(\mathbf{v}_{1}^{\prime},\mathbf{v}_{2}^{\prime
},\mathbf{r}^{\prime};t), \label{49}%
\end{equation}
then it is not difficult to show, using (\ref{47}), that $f_{c}(\mathbf{r}%
,\mathbf{g;}t)$ satisfies the equation
\begin{equation}
\{\mathbf{g\cdot\nabla-}[\mathbf{\nabla}V(\mathbf{r})]\cdot\mathbf{\partial
\}}f_{c}(\mathbf{r},\mathbf{g;}t)=\mathbf{g\cdot\nabla\lbrack}\varphi
_{c}(\mathbf{v}_{1},\mathbf{v}_{2},\mathbf{r};t)] \label{49a}%
\end{equation}
with the limiting condition%
\begin{equation}
\lim_{V(r)\rightarrow0}f_{c}(\mathbf{r},\mathbf{g;}t)=\lim_{V(r)\rightarrow
0}\varphi_{c}(\mathbf{v}_{1},\mathbf{v}_{2},\mathbf{r};t)=0. \label{50}%
\end{equation}
The condition (\ref{50}) follows from the definition (\ref{43'}) for
$\varphi_{c}(\mathbf{v}_{1},\mathbf{v}_{2},\mathbf{r};t)$ and from the natural
observation that the correlation function $g_{2}(\mathbf{r},\mathbf{v}%
_{1},\mathbf{v}_{2};0)=0$ when the interparticle interaction $V(r)$ vanishes.

Now we can write down the function $J_{c}(\mathbf{v}_{1},\mathbf{v}_{2})$,
defining the collision term (\ref{43''''}), as%
\begin{align}
J_{c}(\mathbf{v}_{1},\mathbf{v}_{2})  &  =\int d\mathbf{r}[\mathbf{\nabla
}V(\mathbf{r})]\cdot\partial\mathbf{[}f_{c}\mathbf{(\mathbf{r},\mathbf{g;}%
}t\mathbf{)-}\varphi_{c}\mathbf{(\mathbf{v}_{1},\mathbf{v}_{2},\mathbf{r}%
;}t\mathbf{)]}\nonumber\\
&  =\int d\mathbf{rg\cdot\nabla}f_{c}\mathbf{(\mathbf{r},\mathbf{g;}%
}t\mathbf{)\mathbf{-}}\int d\mathbf{r[}\nabla V\mathbf{(\mathbf{r}%
)]\cdot\partial\varphi}_{c}\mathbf{\mathbf{(\mathbf{v}_{1},\mathbf{v}%
_{2},\mathbf{r};}t\mathbf{)}.} \label{52}%
\end{align}
Here we used (\ref{49a}) and that $\varphi_{c}\mathbf{(\mathbf{v}%
_{1},\mathbf{v}_{2},\mathbf{r};}t\mathbf{)}$ depends on the relative distance
$r=\left\vert \mathbf{r}\right\vert $ , and, therefore, $\int d\mathbf{rg\cdot
\nabla}\varphi\mathbf{(\mathbf{v}_{1},\mathbf{v}_{2},r;}t\mathbf{)=}0$.

It is interesting to note, that the term $\int d\mathbf{r[}\nabla
V\mathbf{(\mathbf{r})]\cdot\partial\varphi}_{c}\mathbf{\mathbf{(\mathbf{v}%
_{1},\mathbf{v}_{2},\mathbf{r};}t\mathbf{)}}$ in (\ref{52}) exactly
compensates the irrelevant correlation term (\ref{43'}), i.e.,
\begin{equation}
n\int d\mathbf{r}\int d\mathbf{v}_{2}L^{\prime}\varphi_{c}(\mathbf{v}%
_{1},\mathbf{v}_{2},\mathbf{r};t)-n\int d\mathbf{r}\int d\mathbf{v}%
_{2}\mathbf{[}\nabla V\mathbf{(\mathbf{r})]\cdot\partial\varphi}%
_{c}\mathbf{\mathbf{(\mathbf{v}_{1},\mathbf{v}_{2},\mathbf{r};}t\mathbf{)=}}0.
\label{52'}%
\end{equation}
Then,%
\begin{equation}
J_{c}(\mathbf{v}_{1},\mathbf{v}_{2};t)=\int d\mathbf{rg\cdot\nabla}%
f_{c}\mathbf{(\mathbf{r},\mathbf{g;}}t\mathbf{)=0.} \label{52''}%
\end{equation}

Thus, at the large timescale, the initial correlations compensate the
irrelevant term (\ref{43'}) and do not contribute to the collision term
$J_{c}(\mathbf{v}_{1},\mathbf{v}_{2};t)$.

Let us consider the last correlation term (\ref{44})
\begin{equation}
n\int d\mathbf{r}\int d\mathbf{v}_{2}%
{\textstyle\int\limits_{0}^{\infty}}
dt_{1}L^{\prime}U_{2}(t_{1})[-\mathbf{g\bullet\nabla]}\varphi_{c}%
(\mathbf{v}_{1},\mathbf{v}_{2},\mathbf{r};t). \label{52'''}%
\end{equation}
Proceeding in the same manner, we introduce the following function%
\begin{align}
f_{c}^{\prime}(\mathbf{r},\mathbf{g;}t)  &  =\varphi_{c}(\mathbf{v}%
_{1},\mathbf{v}_{2},\mathbf{r};t)+\int d\mathbf{r}^{\prime}\int d\mathbf{g}%
^{\prime}Z(\mathbf{r},\mathbf{g};\mathbf{r}^{\prime},\mathbf{g}^{\prime
})\nonumber\\
&  .[-\mathbf{g}^{\prime}\mathbf{\nabla}^{\prime}]\varphi_{c}(\mathbf{v}%
_{1}^{\prime},\mathbf{v}_{2}^{\prime},\mathbf{r}^{\prime};t),\nonumber\\
\lim_{V(r)\rightarrow0}  &  =\lim_{V(r)\rightarrow0}\varphi_{c}(\mathbf{v}%
_{1},\mathbf{v}_{2},\mathbf{r};t)=0. \label{52''''}%
\end{align}
Then, the term (\ref{52'''}) can be rewritten as%
\begin{equation}
J_{c}^{\prime}=n\int d\mathbf{r}\int d\mathbf{v}_{2}L^{\prime}[f_{c}^{\prime
}(\mathbf{r},\mathbf{g;}t)-\varphi_{c}(\mathbf{v}_{1},\mathbf{v}%
_{2},\mathbf{r};t)]. \label{52'''''}%
\end{equation}

On the other hand, we have%
\begin{equation}
\{\mathbf{g\cdot\nabla-}[\mathbf{\nabla}V(\mathbf{r})]\cdot\mathbf{\partial
\}}f_{c}^{\prime}(\mathbf{r},\mathbf{g;}t)=-L^{\prime}\varphi_{c}%
(\mathbf{v}_{1},\mathbf{v}_{2},\mathbf{r};t), \label{52a}%
\end{equation}
i.e., the term (\ref{52'''}) can be written as
\begin{equation}
J_{c}^{\prime}(v_{1},v_{2};t))=n\int d\mathbf{r}\int d\mathbf{v}%
_{2}\mathbf{g\nabla}f_{c}^{\prime}(\mathbf{r},\mathbf{g;}t)=0. \label{52b}%
\end{equation}

Thus, Eq. (\ref{43}) reduces tot%
\begin{equation}
\frac{\partial F(\mathbf{v}_{1};t)}{\partial t}=\int d\mathbf{v}_{2}%
[J(v_{1},v_{2};t)+J_{c}(v_{1},v_{2};t)+J_{c}^{\prime}(v_{1},v_{2};t)]
\label{52d}%
\end{equation}
where $J(v_{1},v_{2};t)$ is defined as%
\begin{equation}
J(v_{1},v_{2};t)=n\int d\mathbf{r}_{2}L_{12}%
{\textstyle\int\limits_{0}^{\infty}}
dt_{1}U_{2}(t_{1})L_{12}F_{1}(\mathbf{p}_{2};0)F_{1}(\mathbf{p}_{1};t),
\label{52e}%
\end{equation}
i.e., at $t\gg t_{cor}$ initial correlations seizes to influence the evolution
of $F(\mathbf{v}_{1};t)$.

\section{The Boltzmann equation}

Let us now introduce the function%
\begin{align}
f(\mathbf{r},\mathbf{g;}t)  &  =F_{1}(\mathbf{p}_{2};0)F_{1}(\mathbf{p}%
_{1};t)\nonumber\\
&  +\int d\mathbf{r}^{\prime}\int d\mathbf{g}^{\prime}Z(\mathbf{r}%
,\mathbf{g};\mathbf{r}^{\prime},\mathbf{g}^{\prime})[\mathbf{\nabla}^{\prime
}V(\mathbf{r}^{\prime})]\cdot\mathbf{\partial}^{\prime}F_{1}(\mathbf{p}%
_{2}^{\prime};0)F_{1}(\mathbf{p}_{1}^{\prime};t). \label{52f}%
\end{align}
It satisfies the following equation with the limiting condition (compare with
(\ref{49a}))%

\begin{align}
\{\mathbf{g\cdot\nabla-}[\mathbf{\nabla}V(\mathbf{r})]\cdot\mathbf{\partial
\}}f(\mathbf{r},\mathbf{g;}t)  &  =0,\nonumber\\
\lim_{V(r)\rightarrow0}f_{c}(\mathbf{r},\mathbf{g;}t)  &  =F_{1}%
(\mathbf{p}_{2};0)F_{1}(\mathbf{p}_{1};t). \label{52g}%
\end{align}
It is note difficult to show from (\ref{52g}) that the function $f(\mathbf{r}%
,\mathbf{g;}t)$ satisfies also the equation
\begin{equation}
f(\mathbf{r},\mathbf{g;}t)=F_{1}(\mathbf{p}_{2};0)F_{1}(\mathbf{p}_{1};t)+\int
d\mathbf{r}^{\prime}G^{0}(\mathbf{r}-\mathbf{r}^{\prime})[\mathbf{\nabla
}^{\prime}V(\mathbf{r}^{\prime})]\cdot\mathbf{\partial}^{\prime}%
f(\mathbf{r}^{\prime},\mathbf{g;}t). \label{52h}%
\end{equation}
Using (\ref{52f}), (\ref{52g}) ), and independence of $F_{1}(\mathbf{p}%
_{2};0)F_{1}(\mathbf{p}_{1};t)$ on $\mathbf{r}$, we can obtain the following
expression for $J(v_{1},v_{2};t)$
\begin{equation}
J(\mathbf{v}_{1},\mathbf{v}_{2};t)=\int d\mathbf{rg\cdot\nabla}%
f\mathbf{(\mathbf{r},\mathbf{g;}}t\mathbf{)} \label{52i}%
\end{equation}

Following \cite{Balescu}, let us select the coordinate system in which axis
$z$ is directed along vector $\mathbf{g}$. Then Green's function
$G^{0}(\mathbf{r}-\mathbf{r}^{\prime})$ has the simple form
\begin{align}
G^{0}(\mathbf{r}-\mathbf{r}^{\prime}) &  =g^{-1}\delta(x-x^{\prime}%
)\delta(y-y^{\prime})\theta(z-z^{\prime}),\nonumber\\
\theta(x) &  =1,x>0,\nonumber\\
\theta(x) &  =0,x<0,\label{53}%
\end{align}
which is in agreement with Eq. (\ref{48}). Inserting (\ref{53}) in
(\ref{52h}), we obtain%
\begin{align}
f(\mathbf{r},\mathbf{g;}t) &  =F_{1}(\mathbf{p}_{2};0)F_{1}(\mathbf{p}_{1};t)+%
{\displaystyle\int\limits_{-\infty}^{z}}
dz^{\prime}g^{-1}[\nabla^{\prime}V(x,y,z^{\prime})]\cdot\mathbf{\partial
}^{\prime}f(x,y,z^{\prime},g\mathbf{;}t)\nonumber\\
&  =F_{1}(\mathbf{p}_{2};0)F_{1}(\mathbf{p}_{1};t)+%
{\displaystyle\int\limits_{-\infty}^{z}}
dz^{\prime}(\frac{\partial}{\partial z^{\prime}})f(x,y,z^{\prime}%
,g\mathbf{;}t)\nonumber\\
&  =F_{1}(\mathbf{p}_{2};0)F_{1}(\mathbf{p}_{1};t)+\Phi(x,y,z,g;t),\nonumber\\
\Phi_{c}(x,y,z,g\mathbf{;}t) &  =%
{\displaystyle\int\limits_{-\infty}^{z}}
dz^{\prime}(\frac{\partial}{\partial z^{\prime}})f_{c}(x,y,z^{\prime
},g\mathbf{;}t),\label{54}%
\end{align}
Finally, introducing (\ref{54}) in (\ref{52i}), we obtain%
\begin{align}
J(\mathbf{v}_{1},\mathbf{v}_{2}) &  =\int d\mathbf{rg\cdot\nabla
}f\mathbf{(\mathbf{r},\mathbf{g;}}t\mathbf{)}\nonumber\\
&  \mathbf{=}%
{\displaystyle\int\limits_{-\infty}^{\infty}}
dzg(\frac{\partial}{\partial z})\Phi(x,y,z,g;t)=g[\Phi(x,y,\infty
,g;t)-\Phi(x,y,-\infty,g;t)]\nonumber\\
&  =g%
{\displaystyle\int\limits_{-\infty}^{\infty}}
dz^{\prime}(\frac{\partial}{\partial z^{\prime}})f(x,y,z^{\prime}%
,g\mathbf{;}t)\nonumber\\
&  =g[f(x,y,+\infty,g\mathbf{;}t)-f(x,y,-\infty,g\mathbf{;}t)].\label{55}%
\end{align}
It is evident from (\ref{54}), that function $f(x,y,-\infty,g\mathbf{;}t)$ is
given by
\begin{equation}
f(\mathbf{v}_{1},\mathbf{v}_{2},x,y,\mathbf{-\infty};t)=F_{1}(\mathbf{v}%
_{2};0)F_{1}(\mathbf{v}_{1};t),\label{56}%
\end{equation}
i.e., by the distribution function of the incoming particles.

The function $f(x,y,+\infty,g\mathbf{;}t)$ is the distribution of outgoing
particles after collision with the relative velocity $\mathbf{g}$ far from the
region of effective interaction. The velocity of these particles
$\mathbf{v}_{1}^{\prime}$ and $\mathbf{v}_{2}^{\prime}$ are such that after
the collision they are exactly equal to $\mathbf{v}_{1}$ and $\mathbf{v}_{2}$,
respectively, and determined by the solution of the two-body problem. But due
to the Liouville theorem, this distribution function is equal to the
distribution function before collision with velocities $\mathbf{v}_{1}%
^{\prime}$, $\mathbf{v}_{2}^{\prime}$, i.e.,
\begin{equation}
f(x,y,+\infty,g\mathbf{;}t)=F_{1}(\mathbf{v}_{2}^{\prime};0)F_{1}%
(\mathbf{v}_{1}^{\prime};t).\nonumber
\end{equation}
Taking into account that in the adopted coordinate system with $\mathbf{g}$
directed along $z$-axis%
\begin{equation}
dxdy=bdbd\varphi, \label{57}%
\end{equation}
where $b$ is the impact parameter and $\varphi$ is the azimuth angle. As a
result, we arrive at the following linear Boltzmann equation (see also
\cite{Catapano 2018})
\begin{align}
\frac{\partial F_{1}(\mathbf{v}_{1};t)}{\partial t}  &  =n\int d\mathbf{v}%
_{2}\int d\varphi dbbg[F_{1}(\mathbf{v}_{2}^{\prime};0)F_{1}(\mathbf{v}%
_{1}^{\prime};t)\nonumber\\
&  -F_{1}(\mathbf{v}_{2};0)F_{1}(\mathbf{v}_{1};t)],t\gg t_{cor}. \label{58}%
\end{align}

Equation (\ref{58}) is obtained with no use of the "molecular chaos"-type
approximations and accounting for initial correlations which disappear at the
large timescale $t\gg t_{cor}$. We see, that on this timescale, when the
influence of initial correlation disappears, the linear for $F_{1}%
(\mathbf{v}_{1};t)$ Eq. (\ref{58}) looks like familiar "nonlinear" equation
for one-particle distribution function but with the distribution function of
the second tagged particle $F_{1}(\mathbf{v}_{2};0)$ taken at the initial
moment of time $t=0$.

In connection with that, let us consider the difference%
\begin{equation}
F_{1}(\mathbf{v}_{2};t)-F_{1}(\mathbf{v}_{2};0)=[\frac{\partial}{\partial
t}F_{1}(\mathbf{v}_{2};t)]_{t=0}t+\ldots\thickapprox\frac{F_{1}(\mathbf{v}%
_{2};t)}{t_{rel}}t+\ldots. \label{58a}%
\end{equation}
i.e., the difference is small at $t\ll t_{rel}$, where $t_{rel}$ is defined by
Eq. (\ref{58}).

Let us consider the time interval
\begin{equation}
t_{cor}\ll t\ll t_{rel} \label{58a'}%
\end{equation}
Then, Eq. (\ref{58}) holds ($t_{cor}\ll t$), the difference (\ref{58a}) is
small ($t\ll t_{rel}$), and we can replace in this equation $F_{1}%
(\mathbf{v}_{2};0)$, $F_{1}(\mathbf{v}_{2}^{\prime};0)$ with $F_{1}%
(\mathbf{v}_{2};t)$, $F_{1}(\mathbf{v}_{2}^{\prime};t)$, respectively. It
means that the second tagged article moves freely before and after collision.

Thus, we obtain the conventional nonlinear Boltzmann equation%
\begin{align}
\frac{\partial F_{1}(\mathbf{v}_{1};t)}{\partial t}  &  =n\int d\mathbf{v}%
_{2}\int d\varphi dbbg[F_{1}(\mathbf{v}_{2}^{\prime};t)F_{1}(\mathbf{v}%
_{1}^{\prime};t)\nonumber\\
&  -F_{1}(\mathbf{v}_{2};t)F_{1}(\mathbf{v}_{1};t)], \label{58b}%
\end{align}
which holds in the time interval (\ref{58a'}) with%
\begin{equation}
t_{cor}\thicksim\frac{b}{v},t_{rel}\thicksim\frac{l}{v}\thicksim\frac
{1}{(nb^{2})v}, \label{58c}%
\end{equation}
where $l$ is the particle mean free path $l\thicksim(nb^{2})^{-1}$.

Therefore, in the framework of our consideration, the Boltzmann equation
(\ref{58b}) holds at any finite $t\gg t_{cor}$, when the mean free path
$l\rightarrow\infty$ ($t_{rel}\rightarrow\infty$).

\section{Summary}

We have introduced a special projection operator (\ref{12}), which being
applied to the Liouville equation for $N$-particle distribution function
$F_{N}(t)$ ($N\gg1$), results in the exact completely closed (homogeneous)
linear equation for an $s$-particle distribution function $F_{s}(t)$ ($s\leq
N$) (\ref{15a}), (\ref{28}) with no inhomogeneous irrelevant initial
correlation term but accounting for initial correlations. This is the main
result of the paper, which solves the problem of the derivation of the closed
evolution equation for $F_{s}(t)$ effectively accounting for initial
correlations with no "molecular chaos" type approximation for an arbitrary
initial state of the system of classical particles, by transfer the initial
correlations to the kernel governing its evolution via the introduced
projection operator, which does not change an arbitrary initial distribution
function $F_{N}(0)$.

For the case of a small particles' density $n$, the obtained equation for
$F_{s}(t)$ was rewritten in the linear approximation in $n$ for the kernel
(see Eqs. (\ref{29}), and (\ref{32})). In particular, the equations for
$F_{1}(t)$ (\ref{34}) and $F_{2}(t)$ (\ref{35}), which are exact in the linear
in $n$ approximation, were obtained. The equation for a one-particle
distribution function was considered in detail and showed that it can be
written differently at different timescales. At $t\thicksim t_{cor}\ll
t_{rel}$, the initial correlations influence the evolution of $F_{1}(t)$ (Eqs.
(\ref{37}), (\ref{39})). At the large timescale $t\gg t_{cor}$, the initial
correlations seize influencing the evolution of $F_{1}(t)$ which is now
described by the linear Boltzmann equation (\ref{58}). In the time interval
$t_{cor}\ll t\ll t_{rel}$, the evolution can be presented by the nonlinear
Boltzmann equation (\ref{58b}) with an accuracy up to the terms of the
$\gamma\frac{t}{t_{rel}}$ order ($\gamma\ll1$ (see (\ref{30}) and $\frac
{t}{t_{rel}}\ll1$) and in the limit $t_{rel}\rightarrow\infty$ (mean free path
$l\rightarrow\infty$), the Boltzmann equation (\ref{58b}) exactly holds for
all finite $t\gg t_{cor}$.

Thus, the Boltzmann equation (\ref{58b}) is derived with no "molecular
chaos"-type assumption in the limit $l\rightarrow\infty$.

\end{document}